\documentstyle{ichep}
\def\simge{\mathrel{%
   \rlap{\raise 0.511ex \hbox{$>$}}{\lower 0.511ex \hbox{$\sim$}}}}
\def\simle{\mathrel{
   \rlap{\raise 0.511ex \hbox{$<$}}{\lower 0.511ex \hbox{$\sim$}}}}

\def\slashchar#1{\setbox0=\hbox{$#1$}           
   \dimen0=\wd0                                 
   \setbox1=\hbox{/} \dimen1=\wd1               
   \ifdim\dimen0>\dimen1                        
      \rlap{\hbox to \dimen0{\hfil/\hfil}}      
      #1                                        
   \else                                        
      \rlap{\hbox to \dimen1{\hfil$#1$\hfil}}   
      /                                         
   \fi}                                         %

\def\ol{\overline}

\def\third{\textstyle{ { 1\over { 3 } }}}

\def\shat{\hat s}
\def\rshat{\sqrt{\shat}}
\def\ecm{\sqrt{s}}
\def\gev{\rm GeV}
\def\tev{\rm TeV}
\def\ts{\thinspace}

\def\ra{\rightarrow}

\def\Lra{\Longrightarrow}
\def\pb{\rm pb}

\def\ifb{{\rm fb}^{-1}}
\def\Mh{M_{\eta_T}}
\def\Mv{M_{V_8}}
\def\tbt{\ol t t}
\def\ttb{t \ol t}

\def\qqb{q \ol q}

\def\ppb{p \ol p}
\def\QbQ{\ol Q Q}

\def\stt{\sigma(\ttb)}
\def\Mtt{{\cal M}_{t \ol t}}
\def\MMtt{\langle \Mtt \rangle}

\def\RMStt{{\langle \Mtt^2 \rangle^{1/2}}}
\def\Deltt{\Delta \Mtt}

\def\jets{{\rm jets}}
\def\pt{p_T}

\def\Dzero{{\rm D}\slashchar{\rm O}}

\begin{document}

\title{Top--Quark Production and Flavor Physics---The Talk$^{\dag}$}

\author{Kenneth Lane$^{\ddag}$}

\affil{Department of Physics,
Boston University,\\
590 Commonwealth Avenue, Boston, MA 02215, USA\\}

\abstract{Because of the top quark's very large mass, about 175~GeV, it now
provides the best window into flavor physics. Thus, pair--production of top
quarks at the Tevatron Collider is the most incisive probe of this physics
until the Large Hadron Collider turns on in the next century. In this talk
I discuss how moments of the $\ttb$ invariant mass distribution can be used
to distinguish among standard and alternative mechanisms of $\ttb$
production.}

\twocolumn[\maketitle]

\fnm{1}{Invited talk given at the 27th International Conference on High
Energy Physics, Glasgow, 20--27th July 1994.}
\fnm{2}{email: lane@buphyc.bu.edu}

\section{Introduction}
The CDF collaboration has reported evidence for top--quark production at
the Tevatron Collider~\cite{cdfpr},\cite{cdfprl}. According to these
papers, the top mass is $m_t = 174\pm 10 \ts ^{+13}_{-12} \,\gev$. The data
in these papers are based on an integrated luminosity of $19.3\,\pb^{-1}$.
When combined with the detector's efficiencies and acceptances, CDF reports
the production cross section $\sigma(\ppb \ra \ttb) = 13.9
\ts^{+6.1}_{-4.8}\,\pb$ at $\sqrt{s} = 1800\,\gev$. The predicted QCD cross
section for $m_t = 174\,\gev$, including next--to--leading--log
corrections~\cite{qcdref} and soft--gluon resummation~\cite{resum} is $\stt
= 5.10^{+0.73}_{-0.43}\,\pb$. This is 2.8~times smaller than the central
value of the measured cross section. The uncertainty in $\alpha_S$
increases the theoretical error $\stt$ to at most~30\%~\cite{ellis}.

The error on the CDF cross section is large, but so is the discrepancy with
QCD. If it holds up, it heralds the long--awaited collapse of the standard
model. In any event, it is clear that the top quark provides a wide--open
window into the world of flavor physics. It is the heaviest elementary
particle we know and, more to the point, the heaviest elementary fermion by
a factor of~40! If the Higgs boson of the minimal one--doublet model
exists, its coupling to the top quark, renormalized at~$m_t=174\,\gev$, is
$\Gamma_t = 2^{3/4}\ts G_F^{1/2}\ts m_t = 1.00$. If charged
scalars---members of Higgs--boson multiplets or technipions---exist, they
couple to top quarks with $O(1)$ strength and they decay as $H^+ \ra t \ol b$.

In this talk, we discuss how moments of the $\ttb$ invariant--mass ($\Mtt$)
distribution may be used to distinguish among competing models of top
production. We point out that, in QCD, the mean and root--mean--square
invariant masses are linear functions of the top--quark mass over the
entire interesting range of $m_t$. Thus, the $\Mtt$ distribution can
provide an {\it independent} determination of the top quark's mass. We
apply this to the existing data~\cite{cdfpr} and find good consistency with
the reported mass. This analysis is made at the simplest theoretical level.
The analysis needs to be carefully done by the CDF and $\Dzero$
collaborations themselves.

The lowest two moments and their variance, $\Deltt = \sqrt{\langle \Mtt^2
\rangle - \MMtt^2}$, can provide valuable discrimination among
top--production models for limited statistics. Examples of this are given
for three models of enhanced $\ttb$--production. The first involves
resonant production of a 400--600~GeV color--octet vector meson
(``coloron''), $V_8$, which is associated with electroweak symmetry
breaking via top--condensation~\cite{topcolor} and which interferes with
QCD production via the process $\qqb \ra V_8 \ra \ttb$~\cite{hp}.  The
second example invokes a color--octet pseudoscalar,
$\eta_T$~\cite{etarefs}. In multiscale models of walking
technicolor~\cite{multi},\cite{wtc}, it is produced strongly in
gluon--gluon fusion and decays mainly to
$\ttb$~\cite{tagteta},\cite{eekleta}. The third model has additional
pair--production of an electroweak--{\it isoscalar}, color--triplet quark,
$t_s$, which is approximately degenerate with the top quark and which,
through mass--mixing, decays as $t_s \ra W^+ b$~\cite{bp}. The agreement
between the directly measured top--mass and that extracted from the $\Mtt$
moments does not yet rule out these new mechanisms of top--quark
production, but it may do so with data from the current Tevatron run.

For reasons of space, our discussions refer entirely to $\ttb$ production
at the Tevatron. The QCD process there is dominated by $\qqb$ annihilation,
as is top production in the $V_8$ model and the isoscalar quark model. As
noted, production in the $\eta_T$ model is dominated by gluon fusion. If
the energy of the Tevatron can be increased to 2~TeV, there will be a 35\%
increase in $\stt$ if the new physics requires $\qqb$ annihilation, but a
65\% increase if the  $\eta_T$ is involved$^{\dag}$\fnm{1}{I thank S.~Parke
for emphasizing this point to me.}. Dramatic differences will occur when
LHC energies are reached. The rate in the $V_8$ model is typically 10--20\%
higher than the standard QCD rate at the LHC (for $|\eta_t| < 1.5$). The
isoscalar quark process remains a few times the standard top--quark rate.
However, the rate in the $\eta_T$ model we consider is 10--15 {\it times}
the standard one, reflecting the importance of gluon fusion for low--$x$
physics. Of course, if there is new physics involved in top--quark
production, its origin should be determined at the Tevatron well before the
LHC turns on.

A more complete version of the results summarized here was submitted to
this conference in Ref.~\cite{klgls} and will appear in
Ref.~\cite{klprtop}. Also discussed there is the angular distribution of
the top quark in $\ttb$ production. Measurements of the angular
distributions will require much larger data sets than will be available in
the next year or two. Thus, to realize the full potential of the top--quark
handle on flavor physics, it is essential that the Tevatron experiments be
able to collect samples as large as 1--10~fb$^{-1}$. Such large data sets
may even help our science avoid Mark Twain's characterization of having
``such wholesale returns of conjecture for such a trifling investment of
fact''~\cite{twain}.

\section{Invariant Mass Distributions}
We calculated the $\Mtt$--distribution expected in QCD
for the Tevatron Collider and top--quark masses in the interesting range
100--200~GeV and found that it is sharply peaked at ${\cal M}_{\rm max}
\simeq 2.1 m_t + 10\,\gev$. As a consequence, low moments of the mass
distribution, the mean and RMS, are nearly linear functions of the
top--quark mass (also see Ref.~\cite{tdr}). We found
\begin{equation}
\eqalign{
\MMtt &= 50.0\,\gev + 2.24 \ts m_t \cr
\RMStt &= 58.4\,\gev + 2.23 \ts m_t \ts. \cr }
\end{equation}
In the range $m_t \simeq$ 140--180~GeV, the dispersion in $\Mtt$ expected
for standard QCD production is $\Deltt =$ 70--75~GeV$^{\ddag}$\fnm{2} {Our
calculations used lowest--order QCD subprocess cross sections and the EHLQ
Set~1 parton distribution functions~\cite{ehlq}. We believe that our
general conclusions will remain true when higher--order corrections are
included. Our $\ttb$ cross sections have been multiplied by a factor of
1.6165. This makes our standard QCD rates as a function of $m_t$ agree to
within a per~cent with the central values quoted in Ref.~\cite{resum} over
the of top masses of interest. The results in Eq.~(1) are accurate so long
as the higher--order corrections are well--represented by a simple
multiplicative factor. Our parton level calculations ignore transverse
motion of the $\ttb$ center--of--mass induced, e.g., by initial--state
radiation. While this effect is not large, it can and should be taken into
account in more detailed simulations.}.

In Ref.~\cite{cdfpr}, the top quark mass was determined from a sample of
seven $W \ra \ell \nu \ts + \ts 4 \ts \ts \jets$ events by making a
constrained best fit to the hypothesis $\ppb \ra \ttb + X$ followed
by $t \ra W^+ b$ with one $W$ decaying leptonically and the other
hadronically. The CDF paper provides the momentum 4--vectors of all
particles in the event. From these,
the central values of of the $\Mtt$ of seven events may be determined.
These gave the following mean and RMS invariant masses and the
corresponding top--masses:
\begin{equation}
\eqalign{
\MMtt &= 439 \pm 11\,\gev \quad \Lra \quad m_t = 173 \pm 5\,\gev \cr
\RMStt &= 443 \pm 11\,\gev \quad \Lra \quad m_t = 172 \pm 5\,\gev \cr
\Deltt &= 59.5\,\gev \ts. \\}
\end{equation}
The errors in Eq.~(2) are estimated by the ``jacknife'' method of computing
the moments omitting one of the seven events. They are {\it not} to be
interpreted as the true experimental errors. Only the CDF group can provide
those.

The results in Eq.~(2) give some confidence that the measured central value
of the top--quark mass, 174~GeV, is accurate. For example, if $m_t =
160\,\gev$ (for which Ref.\cite{resum} predicts $\stt =
8.2^{+1.3}_{-0.8}\,\pb$), we would expect $\MMtt = 409\,\gev$ and $\RMStt =
415\,\gev$, both well below the values determined above. Thus, if something
is going to change in the CDF results from the next large data sample, we
expect it will be the cross section---which would need to be 2--3 times
smaller to agree with the standard model.

\section{Distinguishing Models of Top--Quark Production}
\begin{table*}
\Table{|c|c|c|c|c|c|c|}{
\hline
Model &
$\stt$ &
$\MMtt$ &
$m_t(\MMtt)$ &
$\RMStt$ &
$m_t(\RMStt$) &
$\Deltt$ \\
\hline
LO-QCD (EHLQ1) & 5.13 & 440 & 174 & 447 & 174 & 77 \\
\hline
CDF data   & 13.9 & 439 & 173 & 443 & 172 & 60 \\
\hline
$M_{V_8^-} = 450$& 13.3 & 431 & 170 & 433 & 168 & 46 \\
$M_{V_8^+} = 450$& 11.0 & 465 & 185 & 469 & 184 & 58 \\
\hline
$M_{V_8^-} = 475$& 14.9 & 440 & 174 & 444 & 173 & 53 \\
$M_{V_8^+} = 475$& 10.8 & 482 & 193 & 487 & 192 & 67 \\
\hline
$\Mh = 450$& 13.5 & 432 & 171 & 435 & 169 & 52 \\
$\Mh = 475$& 11.4 & 442 & 175 & 446 & 174 & 55 \\
\hline
$t_s(160)$ $t(175)$& 13.2 & 421 & 166 & 428 & 166 & 77 \\
$t_s(165)$ $t(190)$& 10.0 & 437 & 173 & 444 & 173 & 77 \\}
\caption{$\ppb \ra \ttb$ total cross sections (in pb) at the Tevatron and
their kinematic characteristics for lowest--order QCD, the CDF
data~[1], and the three nonstandard production models with
parameters described in the text.
\label{tabl1}}
\end{table*}

In this section we compute the first two moments and dispersion of $\Mtt$
for various input parameters to three nonstandard models of top production
\cite{hp},\cite{eekleta},\cite{bp}. The lowest order QCD
subprocess cross sections at parton cm~energy $\rshat$ are
\begin{equation}
\eqalign{
{d \hat \sigma(\qqb \ra \ttb) \over {d z}} =&
{\pi \alpha_S^2 \beta \over {9 \shat}} \ts \bigl(2 - \beta^2 + \beta^2
z^2\bigr) \ts, \cr
{d \hat \sigma(gg \ra \ttb) \over {d z}} =&
{\pi \alpha_S^2 \beta \over {6 \shat}} \ts
\biggl\{{1 + \beta^2 z^2 \over {1 - \beta^2 z^2}}\ts
\biggl[1 - {(1-\beta^2)^2 \over{(1-\beta^2 z^2)}}\biggr] \cr
&\ts\ts + {1-\beta^2 \over{1-\beta^2 z^2}} \ts (1 -
\textstyle{{1\over{8}}} \beta^2 + \textstyle{{3\over{8}}} \beta^2 z^2) \cr
&\ts\ts - \textstyle{{9\over{16}}} (1 + \beta^2 z^2)
\biggr\} \ts. \\}
\end{equation}
Here, $z = \cos\theta$ and $\beta = \sqrt{1 - 4 m_t^2/\shat}$. For $\shat
\gg 4 m_t^2$, these cross sections---especially the gluon fusion one---are
forward--backward peaked. But, at the modest $\shat$ at which QCD
production is large, the cross sections are fairly isotropic.

For the ``coloron'' bosons of Ref.~\cite{hp}, we adopted a version
of the model in which $SU(3)_1 \otimes SU(3)_2$ breaks down
to color $SU(3)$, yielding eight massless gluons and equal-mass $V_8$'s.
To study also the angular distributions in $\ttb$ production
(discussed in \cite{klgls},\cite{klprtop}),
we assumed that the $V_8$ couples only
to left--handed quarks with the amplitude
\begin{equation}
\eqalign{
&A(V^a_8(p,\lambda) \ra q(p_1) \ts \ol q(p_2)) =\cr
&\quad g_S \ts \xi_q \ts \epsilon^\mu(p,\lambda) \ts
\ol u_q(p_1) \ts {\lambda_a \over{2}} \ts \gamma_\mu \ts \left({1-\gamma_5
\over{2}}\right)\ts v_q (p_2) \ts. \cr}
\end{equation}
Here, $g_S$ is the QCD coupling and, following Ref.~\cite{hp}, we took
$\xi_t = \xi_b = \pm 1/\xi_q = \sqrt{40/3}$ ($q =u,d,c,s$). For this chiral
coupling, the $\qqb \ra \ttb$ angular distribution in Eq.~(3) is modified
by the addition of
\begin{equation}
\eqalign{
&{d \hat \sigma(\qqb \ra V_8 \ra \ttb) \over {d z}} =
{\pi \alpha_S^2 \beta \over {36 \shat}} \ts (1 + \beta z)^2 \cr
&\qquad\qquad \times
\ts \left\{\ts\biggl|1 + \xi_q \ts\xi_t \ts{\shat \over {\shat - \Mv^2 + i
\sqrt{\shat} \ts \Gamma(V_8)}}\biggr|^2 - 1\right\} \ts.}
\end{equation}
Ignoring all quark masses except $m_t$, the $V_8$ width is
\begin{equation}
\Gamma(V_8) = {\alpha_S \Mv \over {12}}\ts
\biggl\{4\xi_q^2 + \xi_t^2 \left(1 + \beta (1-m_t^2/\Mv^2)\right)\biggr\}
\ts,
\end{equation}
so that $\Gamma(V_8) = 40$ (85)~GeV for $\Mv = 450$ (475)~GeV.
The sign~$\xi_q \xi_t$ of the $V_8$--gluon interference strongly influences
the shape of the $\ttb$ mass distribution.

If there exists a relatively narrow $\eta_T$ decaying predominantly to
$\ttb$, it modifies the gluon fusion cross section in Eq.~(3) by the
addition of the isotropic term~\cite{tagteta},\cite{eekleta}
\begin{equation}
{d \hat \sigma(gg \ra \eta_T \ra \ttb) \over {d z}} = {\pi \over{4}} \ts
{\Gamma(\eta_T \ra gg)\ts \Gamma(\eta_T \ra \tbt) \over {(\shat - \Mh^2)^2
+ \shat \ts \Gamma^2(\eta_T) }} \ts.
\end{equation}
Interference between the $\eta_T$ and QCD gluon--fusion terms is
a small effect and, so, is not displayed here.

So long as the $\eta_T$ may be treated as a pseudo-Goldstone boson, its
decay rates to gluons may be computed from the triangle
anomaly~\cite{etarefs}. We introduce a model--dependent dimensionless
factor $C_q$ in the Yukawa coupling of $\eta_T$ to $\qqb$~\cite{eekleta}.
We expect $|C_q| = O(1)$. Then, the $\eta_T$'s main decay modes are to
two gluons and $\ttb$ and they are given by
\begin{equation}
\eqalign{
\Gamma(\eta_T \ra gg) &= {5 \alpha_S^2 \ts N_{TC}^2 \ts \Mh^3  \over {384 \ts
\pi^3 \ts F_Q^2}} \ts, \cr\cr
\Gamma(\eta_T \ra \tbt) &= {C_t^2 \ts m_t^2 \ts \Mh \ts \beta_q \over {16 \pi
F_Q^2}} \ts.\cr}
\end{equation}
In these expressions, it is assumed that the $\eta_T$ is composed from a
single doublet of techniquarks $Q = (U,D)$ in the ${\bf N_{TC}}$
representation of $SU(N_{TC})$; $F_Q$ is the decay constant of technipions
in the $\QbQ$ sector. We took $N_{TC} = 5$, $F_Q = 30\,\gev$, and $C_t =
-\third$ in calculations. This value of $F_Q$ is typical of the small
techniquark decay constant occuring in multiscale technicolor
models~\cite{multi}. Its smallness is crucial to obtaining a large
$\eta_T$~contribution to $\ttb$ production. The $\eta_T$ width is then
32~GeV for $\Mh = 450\,\gev$, with branching ratios of ${\textstyle{2\over
{ 3 }}}$ and $\third$ to $\ttb$ and $gg$, respectively.

The third model of enhanced top--production we considered is one in which
an electroweak--isoscalar, charge ${\textstyle {2\over { 3 }}}$ quark,
$t_s$, is approximately degenerate with the top--quark and mixes with it so
they have the same decay mode\cite{bp}. If $m_{t_s} = m_t = 174\,\gev$
(say) the expected rate for the top--quark signal is doubled to
$10.2\,\pb$. We illustrate the isoscalar quark model with two cases:
$m_{t_s} = 160$ and $m_t = 175\,\gev$; $m_{t_s} = 165$ and $m_t =
190\,\gev$.

The total $\ttb$ cross sections at the Tevatron and the characteristics
extracted from the $\Mtt$ distributions are displayed in Table~1 for the
CDF data (see Eq.~(2)) and for the above input parameters to the three
nonstandard production models. We stress the following features:

\medskip

\noindent 1.) The CDF data is narrower ($\Deltt = 60\,\gev$) than the QCD
expectation (77~GeV). While this $\Deltt$ is consistent with the resonant
production models, the statistics are so low that we do not consider this
significant. It is a feature worth watching for in future data samples.

\noindent 2.) If $\xi_q \xi_t = -1$ corresponding to the notation $V_8^-$
in the table, the mass distribution is enhanced below the resonance and
depressed above it, and vice-versa for $\xi_q \xi_t = +1$ ($V_8^+$). Thus,
for a given $\Mv$, the extracted value of $m_t$ is somewhat smaller than or
{\it significantly} larger than the directly--measured one, depending on
whether $\xi_q \xi_t = -1$ or $+1$.

\noindent 3.) The $\eta_T$ does not interfere appreciably with the QCD
gluon fusion process. Thus the value of the extracted top mass depends
mainly on $\Mh$. Resonance masses in the range 400--500~GeV return a top
mass close to the directly--measured value.

\noindent 4.) It is easy to double the QCD value of $\stt$ in the isoscalar
quark model: just choose $m_{t_s} = m_t$. But, as could be foreseen, it is
difficult for the isoscalar quark model to give both a 13.9~pb cross
section and an extracted mass close to the directly--measured one. To get a
cross section $\sim~3$~times as large as QCD requires choosing one of the
masses significantly lower than 174~GeV, leading to too small an extracted
value. This model should be the easiest to eliminate with data from the
current Tevatron run.

To sum up: It should be possible to extract valuable information on the
mechanism of $\ttb$ production and, possibly, the physics of flavor, from
even limited statistics on the $\Mtt$ distrbution. We urge the CDF and
$\Dzero$ experimenters to keep this possibility in mind. In the end, of
course, nothing can make up for large data sets, of $O(1-10)\,\ifb$.
{}From these one can carry out incisive studies of the detailed shape of the
mass distribution and of the $\ttb$ angular distribution. At the same
time, one should study subsystem invariant masses to search for alternate
mechanisms of top production---and hints of flavor physics. This promises
a very exciting physics program for the Tevatron Collider.

I am indebted to Alessandra Caner, Sekhar Chivukula, Estia Eichten, John
Huth, Elizabeth Simmons, John Terning and Avi Yagil for many helpful
conversations. This research was supported in part by the Department of
Energy under Contract~No.~DE--FG02--91ER40676.

\end{document}